\documentclass[aps,prc,twocolumn,superscriptaddress,showpacs,amsmath,nofootinbib,floatfix]{revtex4}
\usepackage{graphicx}
\usepackage{bm}

\begin{document}

\title{Primary versus secondary $\bm{\gamma}$ intensities in 
$^{\bm{171}}$Yb$\bm{(n_{\mathrm{th}},\gamma)}$}

\author{A.~Schiller}
\email{schiller@nscl.msu.edu}
\affiliation{National Superconducting Cyclotron Laboratory, Michigan State 
University, East Lansing, Michigan 48824}
\affiliation{Lawrence Livermore National Laboratory, L-414, 7000 East Avenue,
Livermore, California 94551}
\author{A.~Voinov}
\affiliation{Department of Physics and Astronomy, Ohio University, Athens, Ohio
45701}
\affiliation{Frank Laboratory of Neutron Physics, Joint Institute of Nuclear
Research, 141980 Dubna, Moscow region, Russia}
\author{E.~Algin}
\affiliation{Lawrence Livermore National Laboratory, L-414, 7000 East Avenue,
Livermore, California 94551}
\affiliation{North Carolina State University, Raleigh, North Carolina 27695}
\affiliation{Triangle Universities Nuclear Laboratory, Durham, North Carolina
27708}
\affiliation{Department of Physics, Eskisehir Osmangazi University, Meselik, 
Eskisehir, 26480 Turkey}
\author{L.A.~Bernstein}
\affiliation{Lawrence Livermore National Laboratory, L-414, 7000 East Avenue,
Livermore, California 94551}
\author{P.E.~Garrett}
\affiliation{Lawrence Livermore National Laboratory, L-414, 7000 East Avenue,
Livermore, California 94551}
\affiliation{Department of Physics, University of Guelph, Guelph, Ontario N1G 
2W1 Canada}
\author{M.~Guttormsen}
\affiliation{Department of Physics, University of Oslo, N-0316 Oslo, Norway}
\author{R.O.~Nelson}
\affiliation{Los Alamos National Laboratory, MS H855, Bikini Atoll Rd., Los 
Alamos, New Mexico 87545}
\author{J.~Rekstad}
\affiliation{Department of Physics, University of Oslo, N-0316 Oslo, Norway}
\author{S.~Siem}
\affiliation{Department of Physics, University of Oslo, N-0316 Oslo, Norway}

\begin{abstract}
The two published literature values [Greenwood \sl et al.\rm, Nucl.\ Phys.\ \bf
A252\rm, 260 (1975) and Gelletly \sl et al.\rm, J. Phys.\ G \bf 11\rm, 1055 
(1985)] for absolute primary $\gamma$ intensities following thermal neutron 
capture of $^{171}$Yb differ in average by a factor of three. We have resolved 
this conflict in favor of Greenwood \sl et al.\ \rm by a measurement of primary
versus secondary intensities.
\end{abstract}


\maketitle

Primary, high-energetic ($\agt 3.6$~MeV) and secondary, low-energetic, 
($\alt 2.4$~MeV) $\gamma$-ray intensities from the 
$^{171}$Yb$(n_{\mathrm{th}},\gamma)$ reaction have been published by two groups
in the past \cite{GR75,GL85}\@. While they agree on absolute secondary 
intensities and relative primary intensities, values for absolute primary 
intensities differ in average by a factor of $\sim 3$\@. The potential 
problem with \cite{GR75} is that the authors of this study only measured 
relative primary intensities; these relative intensities were then brought to 
an absolute scale by normalizing to a previous measurement on absolute primary 
intensities \cite{RH69,OR70}\@. The potential problem with \cite{GL85} is that 
primary and secondary $\gamma$ rays were measured with two distinct detector 
systems, hence, a systematic error in the absolute efficiency calibration of 
the high-energy $\gamma$ detector could have resulted in wrong primary 
intensities. The ENSDF compilation \cite{ENSDF} gives a slight preference for 
the primary intensities of \cite{GL85}, however, it also mentions that the 
large disagreement between the two groups is not understood. 

We have performed measurements on the $^{171}$Yb$(n_{\mathrm{th}},\gamma)$ 
reaction at the Los Alamos Neutron Science Center (LANSCE) with two 
large-volume ($\sim 80$\%) Ge(HP) detectors placed at beam height at an angle 
of $\sim 110^\circ$ with the beam, and one actively shielded $\sim 200$\% 
clover detector placed vertically over the target. All three detectors were at 
a distance of $\sim 12$~cm from the target. The target consisted of $\sim 1$~g 
of enriched ($\agt 95\%$), dry $^{171}$Yb$_2$O$_3$ powder encapsulated in a 
glass ampule which was mounted in an evacuated beam tube and irradiated by 
collimated neutrons with a time-averaged flux of $\sim 4\times 10^4$ n/cm$^2$s 
over $\sim 150$~h. The relative detector efficiencies from 1--9~MeV were 
determined by the $^{35}$Cl$(n,\gamma)^{36}$Cl reaction and its known $\gamma$ 
intensities \cite{CB96}\@. The experiment was divided into two runs. During the
first run, the target ampule hung roughly diagonally, having a $\sim 45^\circ$
angle with the vertical and a $90^\circ$ angle with the beam. During the second
run, the target ampule was mounted parallel to the beam. Two efficiency 
calibrations, running $\sim 12$~h each, were performed with corresponding 
orientations of ampules containing $\sim 1$g of NaCl and $\sim 1$g of dried 
AlCl$_3$, respectively. 

\begin{figure*}[tbh!]
\includegraphics[totalheight=9.2cm]{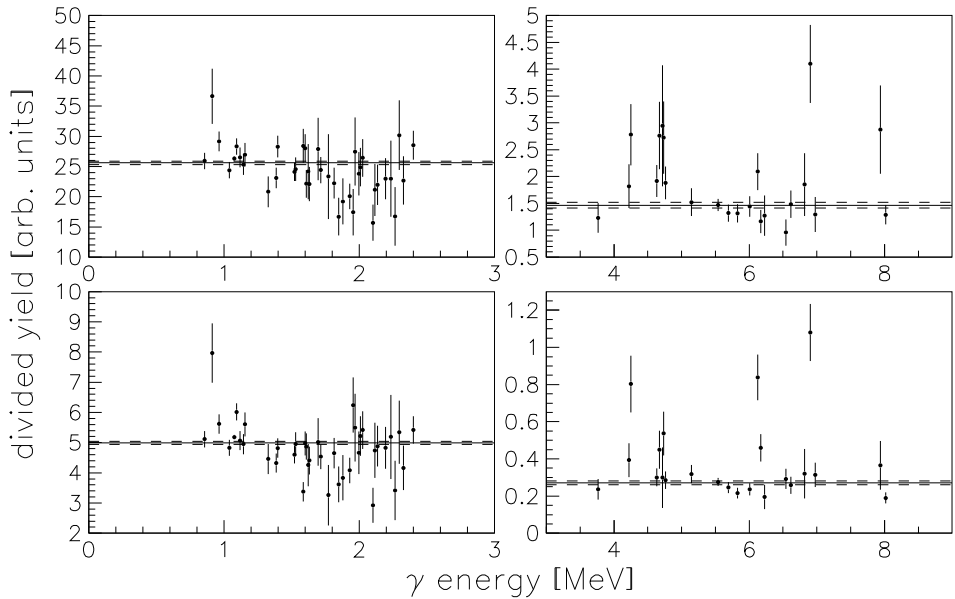}
\caption{Divided yields of secondary (left panels) and primary (right panels) 
$\gamma$ rays. The lines show average values of the divided yields (solid 
lines) and their error bands (dashed lines)\@. Upper and lower panels 
correspond to diagonal and parallel mounts of the target ampules, respectively,
and different chlorine salts used for the efficiency calibration (see text)\@.}
\label{fig:result}
\end{figure*}

The main purpose of the experiment was to measure $\gamma$-$\gamma$ 
coincidences, however, downscaled (by a factor of 100) singles $\gamma$ rays 
were also recorded. More details of the main experiment are given in 
\cite{VS04,SV06}\@. While the experiment was not aimed at measuring absolute 
intensities and it lacked the sensitivity to determine relative primary or 
secondary intensities better than in \cite{GR75,GL85}, the consistent 
efficiency calibration of our detectors over a range of 8~MeV gave us the 
opportunity to investigate the main point of disagreement between \cite{GR75} 
and \cite{GL85} by measuring primary versus secondary intensities. For this 
reason, we have divided summed peak areas from all three detectors for 
individual primary and secondary $\gamma$ rays by the summed relative detector 
efficiencies and relative intensities from \cite{ENSDF} to produce divided 
yields (see Fig.\ \ref{fig:result})\@. Divided yields are then averaged for all
primary and all secondary $\gamma$ rays to produce average divided yields.

The ratio of average divided yields between secondary and primary $\gamma$ rays
is 17.5(7) and 18.4(7) for the two runs, respectively, or 17.9(5) combined. 
This ratio has to be compared to the ratio of normalization factors given by 
\cite{GR75,GL85} and quoted in \cite{ENSDF} to bring the relative secondary and
primary intensities to an absolute scale. These ratios are 
$0.063/0.00107\approx 58.9\pm50\%$ systematic uncertainty for \cite{GL85} and 
$0.054/0.00287\approx 18.8(4)$ for \cite{GR75}\@. For the latter, the error was
calculated assuming that the quoted factors have uncertainties in the order of 
their least significant digits. Hence, we can conclude that \cite{GR75} gives 
correct values of primary intensities (or rather a correct conversion factor to
bring relative primary intensities to an absolute scale), while \cite{GL85} 
does not. To be more precise, we have only shown that \cite{GR75} has the 
correct ratio of absolute primary versus secondary intensities; with our 
experiment we cannot rule out the possibility that both primary and secondary 
intensities from \cite{GR75} might be off by the same factor. However, since 
there does not seem to be a strong disagreement between different measurements 
of absolute secondary intensities, we do not expect this to be a serious 
problem. The $1.5\sigma$ disagreement between our work and \cite{GR75} 
(provided the error estimate for the latter is correct) might be attributed to 
systematic uncertainties in our efficiency calibration due to the 
parameterization of the efficiency curve, absorption in the target, and other 
minor corrections which were all neglected here. 

This work has benefited from the use of the Los Alamos Neutron Science Center 
at the Los Alamos National Laboratory. This facility is funded by the U.S. 
Department of Energy under Contract W-7405-ENG-36\@. Part of this work was 
performed under the auspices of the U.S. Department of Energy by the University
of California, Lawrence Livermore National Laboratory under Contract 
W-7405-ENG-48, and Los Alamos National Laboratory under Contract 
W-7405-ENG-36\@. Financial support from the Norwegian Research Council (NFR) is
gratefully acknowledged. A.V. acknowledges support from a NATO Science 
Fellowship under project number 150027/432 and from the National Nuclear 
Security Administration under the Stewardship Science Academic Alliances 
program through U.S. Department of Energy Research Grant No.\ 
DE-FG03-03-NA00074\@. E.A. acknowledges support by U.S. Department of Energy 
Grant No.\ DE-FG02-97-ER41042\@. We thank Gail F. Eaton for making the targets.

\end{document}